\documentclass[sigconf]{acmart}
\renewcommand\footnotetextcopyrightpermission[1]{} 

\makeatletter
\renewcommand\@formatdoi[1]{\ignorespaces}
\makeatother

\acmConference[Fragile Earth workshop]{KDD 2022}{August 15, 2022}{Washington D.C, USA}
\acmYear{2022}
\copyrightyear{2022}

\AtBeginDocument{%
  \providecommand\BibTeX{{%
    \normalfont B\kern-0.5em{\scshape i\kern-0.25em b}\kern-0.8em\TeX}}}

\copyrightyear{none}
\acmYear{none}
\acmDOI{none}

\setcopyright{none}

\begin{document}

\title{CIMF: Climate impact modelling framework}


\author{Blair Edwards}
\affiliation{%
 \institution{IBM Research Europe}
 \streetaddress{Hartree Centre}
 \city{Daresbury}
 \country{United Kingdom}}

\author{Paolo Fraccaro}
\affiliation{%
 \institution{IBM Research Europe}
 \streetaddress{Hartree Centre}
 \city{Daresbury}
 \country{United Kingdom}}

\author{Nikola Stoyanov}
\affiliation{%
 \institution{IBM Research Europe}
 \streetaddress{Hartree Centre}
 \city{Daresbury}
 \country{United Kingdom}}
 
\author{Nelson Bore}
\affiliation{%
 \institution{IBM Research}
 \streetaddress{Johannesburg}
 \city{Johannesburg}
 \country{South Africa}}
 
\author{Julian Kuehnert}
\affiliation{%
 \institution{IBM Research}
 \streetaddress{Hartree Centre}
 \city{Nairobi}
 \country{Kenya}}

 \author{Kommy Weldemariam}
\affiliation{%
 \institution{IBM Research}
 \streetaddress{Yorktown Heights}
 \city{New York}
 \country{USA}}

\author{Anne Jones}
\affiliation{%
 \institution{IBM Research Europe}
 \streetaddress{Hartree Centre}
 \city{Daresbury}
 \country{United Kingdom}}


\renewcommand{\shortauthors}{B. Edwards, et al.}

\begin{abstract}
  The application of models to assess the risk of the physical impacts of weather and climate and their subsequent consequences for society and business is of the utmost importance in our changing climate.  The operation of such models is historically bespoke and constrained to specific compute infrastructure, driving datasets and predefined configurations.  These constraints introduce challenges with scaling model runs and putting the models in the hands of interested users.
  Here we present a cloud-based modular framework for the deployment and operation of geospatial models, initially applied to climate impacts.  The Climate Impact Modelling Frameworks (CIMF) enables the deployment of modular workflows in a dynamic and flexible manner.  Users can specify workflow components in a streamlined manner, these components can then be easily organised into different configurations to assess risk in different ways and at different scales.  This also enables different models (physical simulation or machine learning models) and workflows to be connected to produce combined risk assessment.  Flood modelling is used as an end-to-end example to demonstrate the operation of CIMF.
\end{abstract}

\keywords{Climate, Models, Impact Modeling, Cloud}

\maketitle

\section{Introduction}

Computational models of a variety of forms play a central role in global efforts to understand, predict and mitigate the impacts of climate change. Modelling of the climate system is coordinated by the climate science community in experiments such as CMIP6 (Eyring et al., 2016). However, modelling of the consequences, or impacts, of climate change to human and natural systems is a more diverse activity. It encompasses a wide range of scientific and technical domains, and a huge variety of stakeholders, from policy makers at the country level, to private sector organisations who need to build in future climate change to their planning and operational processes. Consequently, climate impact modelling software tools are often bespoke, developed for a particular model or use-case, and require substantial effort to implement curation of data, deployment of models and post-processing of outputs. Nevertheless, there are numerous commonalities across climate impact modelling activities, for example, the common requirement for driving models with dynamic climate data and static geospatial data, the need to calibrate and validate models, to scale them efficiently in time and space, and to provide not only predictions but quantification of uncertainties in those predictions. 

In this paper, we present the Climate Impact Modelling Frameworks (CIMF) designed to accelerate the process of climate impact modelling by enabling common tasks through the deployment of composable and reusable modules. CIMF flexibly connects models with geospatial and temporal driving data, facilitates local AI-enhanced model tuning/calibration and validation, and enables models to be deployed quickly and at scale in any location, with customisable workflows to enable uncertainty quantification and flexible summary metrics support tailored use-case-specific output. In addition to reducing the human effort required to configure impact modelling pipelines, CIMF further enhances the process of impact modelling by improving accuracy, reducing computational cost, and facilitating sophisticated uncertainty quantification.

In the following sections, we will lay out the motivation behind the development of the framework and explain the benefits over existing tools.  We will provide motivational use-cases before describing the design and implementation of the framework itself and demonstrate its application to pluvial flood modelling.

\section{Background and Motivation}
\subsection{Climate Risk and Impact}
The defacto standard in climate impact assessment is to take a risk-based approach~\cite{IPCCsummary2022}. Climate risk is the potential for adverse consequences of climate change that affect human or natural systems. Climate change adaptation is then a risk management activity, consisting of developing plans, actions, strategies, or policies to reduce the likelihood and/or magnitude of these adverse consequences. Essential to inform this activity is quantitative information about current and future climate risks generated by climate impact models. These models, or more typically, multi-model pipelines, translate weather and climate variables (e.g. rainfall, temperature, sea level) to climate hazards (e.g. flooding, wind storms, drought) and their consequences (e.g. disruptions to energy or agricultural productivity). Calculation of overall climate risk is carried out by combining three components: hazard, exposure and vulnerability. Climate hazards are the physical events or trends such as extreme rainfall, floods, wildfire, and drought which may have negative consequences. These are arguably the most challenging to support technically, since computational hazard models have a wide variety of forms (simulation/physics-based, data-driven/statistical or rules-based/empirical), and vary hugely in their complexity, configuration requirements, computational intensity, sophistication, and legacy. The final two components are simpler: exposure is the presence of populations, infrastructure or assets in places or settings that could be affected, and can consist of datasets specifying the location and extent/magnitude of these components. Vulnerability is the propensity of exposed elements to be adversely affected, typically modelled using loss or damage functions derived from observational data. 

Climate impact assessment at a global level is coordinated by the Inter-governmental Panel on Climate Change (IPCC) reporting cycle, which collates and summarises scientific consensus on impacts approximately every seven years, with the most recent report, AR6, released in 2021/2022. In AR6, climate risks are projected for near-term (2021-2040), mid term (2041-2060) and long term (2081-2100) at global warming levels of 1.5, 2, 3 and 4$^{\circ}$C above the pre-industrial baseline (1850-1900). Impacts are broken down by affected system/infrastructure (e.g. terrestrial, freshwater, or ocean ecosystems, infectious disease, mental health, heat and cities, settlements and infrastructure) and geographic region~\cite{IPCCfull2022}.

Climate risk quantification and impact modelling studies are carried out both by the scientific community (as collated by IPCC reports) and to address the specific needs of organisations and governments to plan for climate change adaptation. These activities can range from assessment of specific hazards at the country scale (e.g.~\cite{Sayers2020}), to climate change impact assessment for an entire sector or natural system, either globally or regionally (e.g.~\cite{hummel_interacting_2020};~\cite{borrelli_land_2020}), to local government studies (e.g.~\cite{ngai2020},~\cite{bristolone2020}), and finally to private sector impact assessments for organisations own assessment of their exposure to climate risk (\cite{goldstein_private_2019}, \cite{pinchot_assessing_2021}).

Climate impact modelling poses a number of technical challenges. These include the need to: 1) Drive models with spatiotemporal datasets from a range of sources, often exacerbated by the scale of many such datasets, which can easily multiply with increasing spatial and temporal resolution; 2) Accommodate models (particularly hazard models) of a variety of forms, as described above; 3) Support a multitude of use-case specific scenarios and questions; 4) Quantify the uncertainties introduced with each model and dataset. One consequence of the complexity and lack of scalability in the impact modelling process is that climate change impact studies are laborious and time consuming, with timescales of years involved to integrate models and data across different scientific disciplines. Often, impact modelling results are out of date by the time they are released, because by that time the datasets used have been updated.  For example, seven different studies are cited by the 2022 IPCC AR6 report on projected future damages from flooding \cite{IPCCfull2022}, and all seven were based on projections from a subset of the CMIP5 climate model projections released in 2013, which have now been superseded by the CMIP6 dataset released in early 2021.

\subsection{Climate Impact Modelling Tools and Technologies}

Numerous modelling tools and technologies have been developed to support climate impact assessment and modelling. In the simplest case, web-based tools often allow users limited functionality: typically the exploration of pre-computed climate impact modelling results. For example, in the UK, standalone online tools enable access to climate impact studies on erosion and flood risk (\url{https://arcoes-dst.liverpool.ac.uk/}), and crop yield (\url{https://cropnet-demonstrator.datalabs.ceh.ac.uk/}), whereas simple climate risk indicators in the form of temperature and rainfall changes can be explored interactively on the Climate Risk Indicators website (\url{https://uk-cri.org/}).

In the academic research community, tools to enable impact modelling are often bespoke and domain-specific, addressing a limited number of impacts or considering only specific case studies. For example, \cite{masia_modelling_2021} deployed an evapotranspiration model to quantify the impact of climate change on crop water resource requirements. The model was integrated into a GIS spatial platform to couple it with dynamic climate data and static geospatial data for the Mediterranean region. A limitation of the work was a lack of uncertainty quantification since only a single impact model and a single climate model were used in the analysis. This could be remedied by taking a more generic approach.

Some of the most sophisticated modelling tools are those that have been developed to support end-to-end impact modelling in the (re)insurance sector. Examples include CLIMADA \cite{bresch_climada_2021}, and the Oasis Loss Modelling Framework (\url{https://oasislmf.org/}). Both tools are open source, and offer the functionality to combine hazard, exposure and vulnerability to calculate overall risk or expected loss. CLIMADA offers built-in global datasets for key hazards (tropical cyclones, riverine flood, agricultural drought, European winter storms), under different climate change scenarios, along with impact functions. However, the framework requires users to supply their own hazard data if other hazards are needed, or if more sophisticated modelling is required, and does not facilitate running hazard models themselves. 

The Oasis Loss Modelling Framework is an open-source loss modelling software framework for developing, deploying and executing catastrophe models (which include climate-related catastrophes). The framework, contains a multitude of components (\url{https://github.com/OasisLMF}), including support for running hazard models – either supplied by users, contributed by the community or licensed via the Oasis hub (\url{https://oasishub.co/}), including cloud deployment at scale. However, the framework does require users to prepare the geospatial datasets and other input files required by models. For example, \cite{KNOS2022102679} used Oasis to model losses from flash flooding in a Swedish city, employing a simple rainfall intensity metric for flood hazard. Geospatial datasets were pre-processed (e.g. aggregation and interpolation of rainfall data, and alignment of building locations, rainfall events and insurance claim data) separately in ESRI ArcGIS to prepare files for the hazard model which was deployed in the Oasis ktools calculation engine. One reason the authors noted for not utilising a full hydrological/hydraulic model for their analysis, was the laborious process of setting up a complex model for multiple locations.

\section{Use case scenario}

\begin{figure}[t!]
    \centering
    \includegraphics[width=0.5\textwidth]{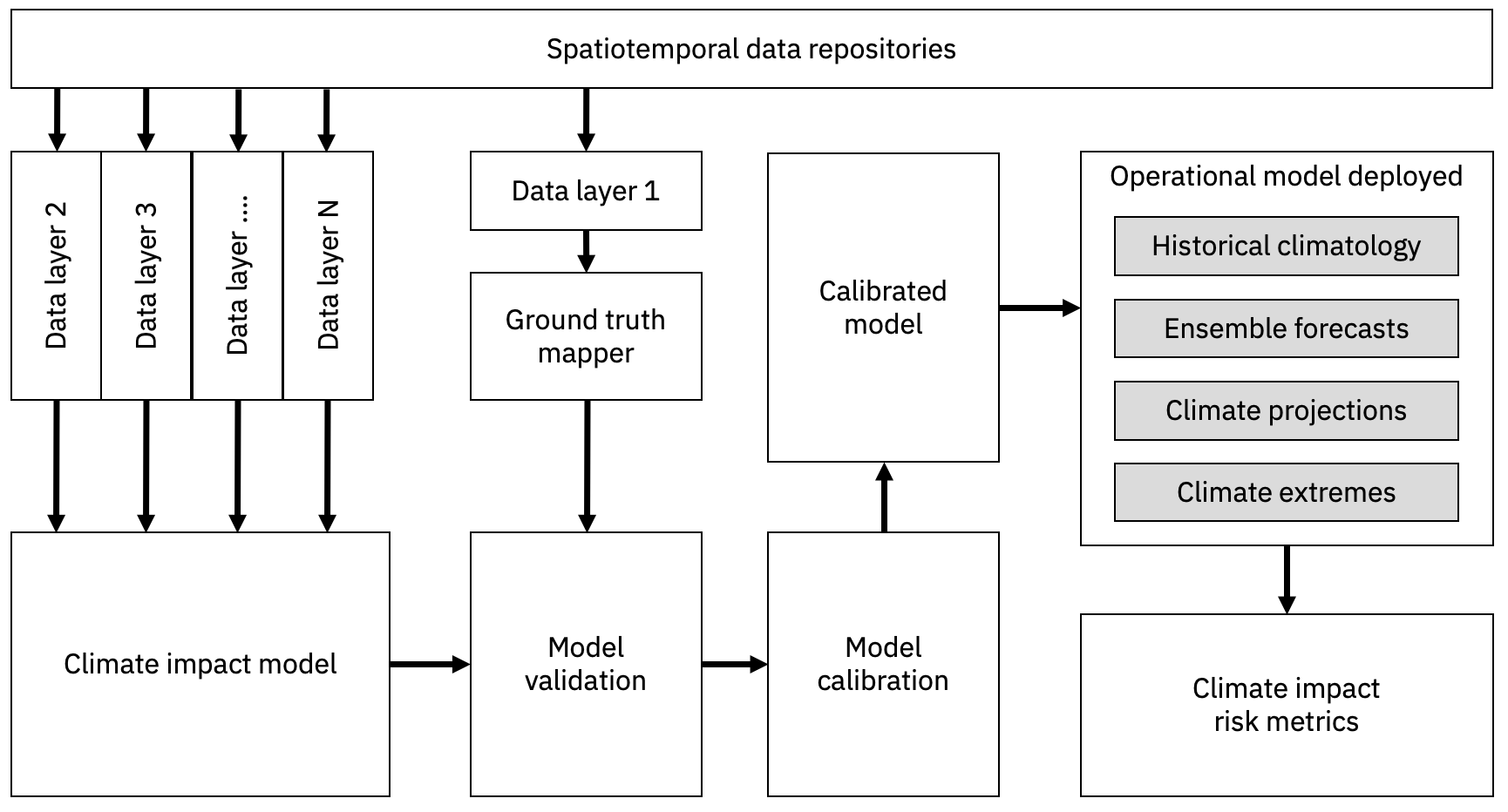}
    \caption{Outline of a typical impact modelling process.  A climate impact model is driven by a range of data layers, validated and calibrated, before being used to calculate risk metrics.}
    \label{fig:impattern}
\end{figure}

Figure~\ref{fig:impattern} shows a typical impact modelling journey.  CIMF enables users to deploy on-boarded models and datasets to address a range of questions and tasks.  With a climate impact model onboarded, users can run that model fed with historical data to evaluate the model performance, validate it (using an automated process) and use this validation to calibrate the models (using an automated service to tune model parameters).  Such validation and calibration capabilities are important to interpret and demonstrate the value of the impact model for a particular location and type of event.  

The calibrated model can subsequently be used to assess historical risk by running a climatology (an ensemble of inputs from a previous years), used to generate risk forecasts on different time horizons (e.g., weather timescale, sub-seasonal or seasonal), climate change projections or for extreme events.  To extract risk metrics from ensembles of model runs, CIMF has the capability to dynamically calculate risk metrics through a simple API call, based on ensemble outputs.  In addition to calculating the probability of individual impacts from single models, multiple impact models can be run with the output compounded in an appropriate manner.  Models can also be chained together to assess the overall impact, combining hazard forecasts or extreme event scenarios, with modules to assess damage to land, property, or people, and even subsequently to the financial cost of such damage.

\begin{figure}
    \centering
    \includegraphics[width=0.5\textwidth]{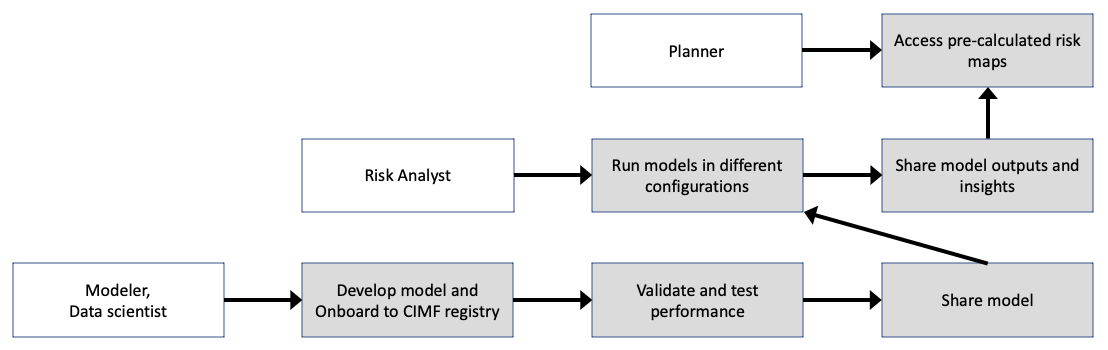}
    \caption{Example user personas and interactions with CIMF.}
    \label{fig:cimf-users}
\end{figure}

One of the key challenges we address with CIMF is flexibility.  This is in terms of deployment configuration, types of models and datasets, but also in terms of serving different user groups.  There are different touch points with geospatial modelling and analytics, for users with different levels of technical and domain expertise.  Due to the way CIMF is designed, it allows the full spectrum of user types to be catered for, as shown in Fig~\ref{fig:cimf-users} and described below. 

Data scientists and modellers from different domains (such as hydrology, agriculture or weather) who develop physical simulation or machine learning models have the ability to onboard their own models into the framework.  This makes it easier for them to scale (spatially and in time), compose workflow components in a simple and intuitive manner, easily parse different datasets to their models, as well as take advantage of other apps/services within the framework (for example model validation or calibration).  The framework gives these advanced users the ability to iterate on models and drive models with the latest/new datasets.  In addition to the enhancements to their own modelling work, the framework also makes it easy for them to share their model with other users and to combine/chain their model with others already onboarded for a more complete impact assessment (i.e. for vulnerability).

There are users who may have domain expertise (e.g., risk analysts), and desire the ability to run models for different spatial domains, on different time horizons, with different inputs datasets or with control of model parameters, but they don't necessarily have the technical expertise to install, code or configure models themselves.  Such users can interact with CIMF through the user-friendly API or a graphical user interface, which abstracts away the details of running the model and allows them to specify the workflow options in a simple JSON (see Fig.~\ref{fig:cimf-payload}), or web form.  They can then access the model outputs either through the API, a dashboard, or through CIMF SDK.

The final user category is those who only require access to risk maps generated by models in the framework.  These can be generated by expert users, on a regular, scheduled basis as appropriate, and shared through a geospatial data portal, such as the IBM Environmental Intelligence Suite.

In addition, to these user categories, the ability to onboard and share models (as a "digital-twin" enabler) between different users, means CIMF reduces the barriers to collaboration between different user groups and scientific domains, for example between climate scientists (developing weather and climate models), impact modellers (studying flooding), people studying economic impacts (modelling damage or financial impacts) and planners or policy makers.

\section{Framework Design, Implementation and Deployment}
The design principles adopted for the development of CIMF were modularity, flexibility, scalability and ease of use.  The framework is built around two key concepts: modules and workflows.  

Modules are blocks of code, which could include anything from short standalone scripts to large complex models with many libraries and dependencies.  These modules are the atomic blocks used to compose workflows.  Modules within CIMF include actions like pulling data from different sources, transforming data, running simulation or machine learning models, processing model outputs or pushing model outputs to external repositories running on the user's target computing environment (e.g., Cloud, On-prem, HPC).  

\begin{figure}
    \centering
    \includegraphics[width=0.5\textwidth]{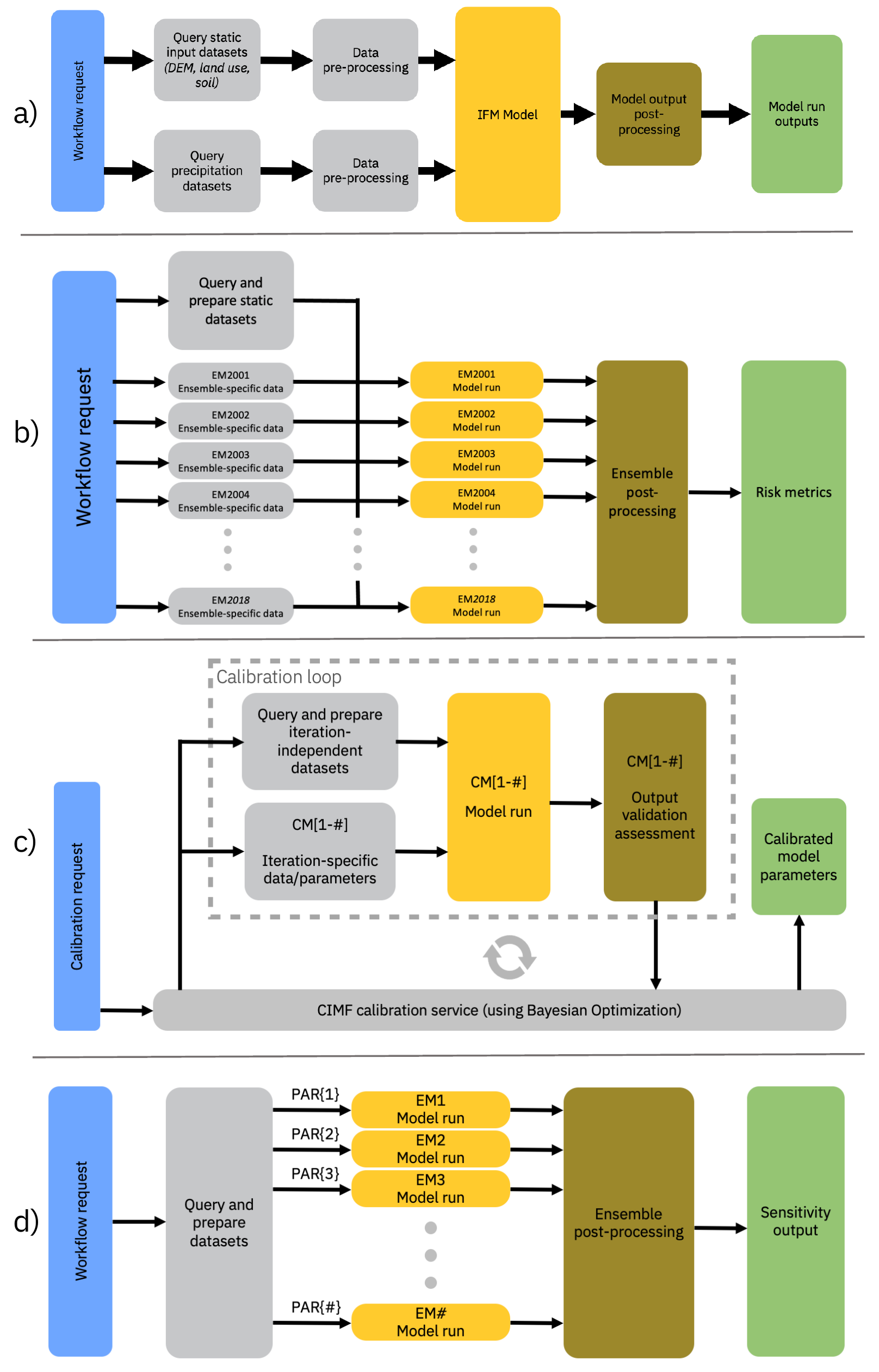}
    \caption{CIMF workflows - example workflow flavours, achieving different risk measurements using a single set of modules.}
    \label{fig:cimf-workflows}
\end{figure}

Workflows are the pipelines of these modules that pass data between them to achieve a specific task, for example to run a flood model.  This  involves querying data, preparing it for the model and then executing the model.  By modularizing workflows into containerized blocks, we are then able to deploy them in a flexible manner, servicing a range of different workflow profiles, without additional coding.  Such workflows can be serial, but can also include branching and even iteration.  When a workflow is submitted, the DAG is generated based on a template and user-selected configuration options, and then executed by a workflow orchestration engine. Figure \ref{fig:cimf-workflows} shows examples of different workflow flavours for a particular simulation flood model, where the same set of modules are used to achieve different goals.  They show a single simulation of a historical flood event, an ensemble of precipitation inputs (for example a historical climatology or seasonal forecast), a ensemble of input parameters (for example exploring model sensitivity to different inputs) or iterative calibration of model parameters.  This is particularly useful when sharing workflows and/or workflow components with other users.  In addition to this, many modules are general purpose and can be reused by a range of workflows for different types of models, reducing the effort of individual modellers in bringing their models to the framework.  

The additional motivation for modularizing workflow steps into containers is around portability and consolidation of workflow patterns.  In order to enable flexibility of deployment for modules, between cloud-enabled clusters, local compute or HPC environments, we decided to utilise S3-style cloud-based object storage as the data interchange layer between workflow steps.  This allows modules to access workflow data from any location, provides persistent storage and easy integration with existing external tools and services.  When a user requests a new workflow run, the framework creates an S3 bucket (for the exclusive use of that workflow instance), and inserts the required input or configuration files.  As the workflow steps (i.e. modules) are executed they will pull and push data to/from the bucket. 

\begin{figure}
    \centering
    \includegraphics[width=0.4\textwidth]{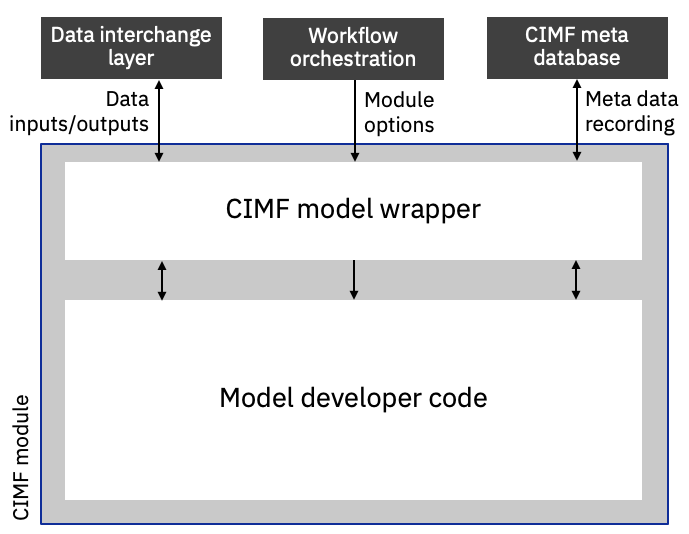}
    \caption{Schematic of the construction of a CIMF module, showing how the model wrapper handles interface between the developers code and the rest of the framework.}
    \label{fig:cimf-module}
\end{figure}

In order to facilitate data transfer and other interactions with the framework, CIMF modules are built including a model wrapper.  Figure \ref{fig:cimf-module} shows a schematic of a CIMF module, with the CIMF model wrapper handling interaction between the framework and the user code, as well as data transfer.  Modules can currently be on-boarded in two ways.  For simple modules, a developer can provide a git repository containing the code in executable form or as a script (i.e. bash or python) and the requirements (e.g. a \emph{requirements.txt} for a Python code) for the code to run.  In this instance, the framework will build a module using a minimal base image including the CIMF model wrapper code and installing the user code inside, before pushing to the container registry.  In the case of more complex codes, a developer can carry out the containerization step themselves, allowing the inclusion and configuration of libraries and dependencies, and the framework will then build the module by installing the CIMF model wrapper into the users base image. These two approaches allow flexibility for users to onboard a wide range of codes with minimal additional effort.  The module on-boarding process utilises a CI/CD backend through Tekton, triggered by a user API call providing a payload which includes the meta-data for the module.  Required meta-data for on-boarding includes a module name, the url for the code in Git or name of the user-created base image, the command needed to run the user code (at run time) and a default definition of the files which are needed by the module to run and those created by the module (i.e. inputs and outputs).  The latter two inform the pull and push of data to/from S3 when the module is executed in a workflow.

\begin{figure}
    \centering
    \includegraphics[width=0.45\textwidth]{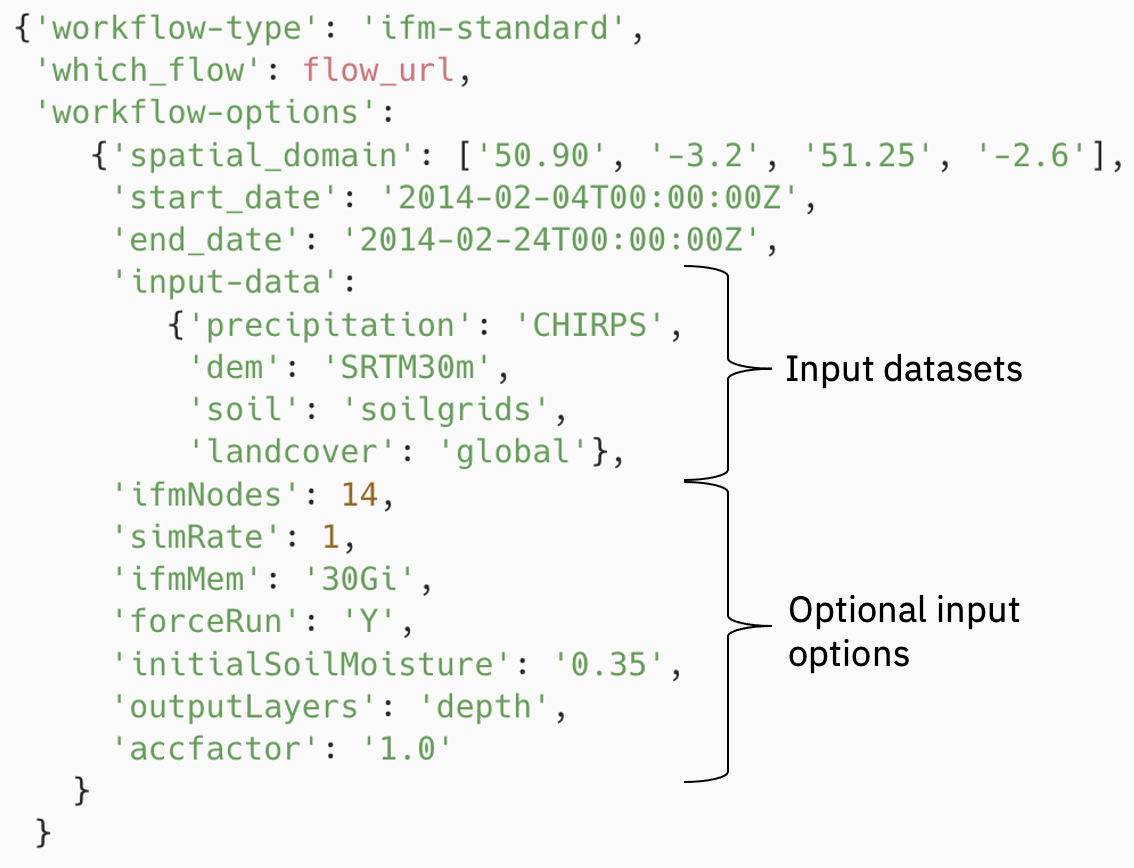}
    \caption{Example API payload required to run a flood model through CIMF, including the workflow type, spatial and temporal domain and workflow options.}
    \label{fig:cimf-payload}
\end{figure}

Workflows are currently defined as a set of templates which are created by the developer at on-boarding time.  These templates describe the modules (i.e. image name and tag) which are to be executed, the dependencies between them, the resources to be allocated to that pod and defaults of parameters which can be passed to the modules at runtime.  These workflow templates are stored in GitHub which provides versioning and tracking.  When a user requests a workflow to run they pass options such as those in Figure~\ref{fig:cimf-payload}.  These options are related to domain specific users inputs, such as the spatial domain of interest, the time horizon over which to run the model and the datasets to use to drive the workflow.  The API translates these user-friendly, domain-relevant inputs into input parameters for the workflow template.  In addition to directly passing values to the workflow engine, more complex inputs can be passed into the workflow through configuration or input files in S3.  An example of such an input would be the payload required to query an external data-source, or the input parameters to a model.  In the case of passing input/configuration files through S3, we append a hash of the file contents to the file name, which allows the framework to check for matching contents from the filename stored in the meta-data.  By abstracting the detailed configuration options away from users (whilst still enabling it for those who desire), we make it much easier for users to use models.

When a user requests a workflow to be run through the CIMF API, an entry is recorded in the Previous Workflow Catalogue (PWC), this includes the user payload, the payload the API passed to the workflow engine and the version of the workflow template used (recorded as the git hash).  This provides traceability and reproducibility of model runs, important for end-users of the model outputs, such as policy-makers.  In future, this could potentially be integrated with blockchain technology to provide provenance of workflow executions.  In addition to providing a record, the PWC can also we used to check if modules or workflows have been run previously allowing the outputs to be reused if appropriate.  This can potential save a significant amount of compute power and user time.  This check is done my stepping through the workflow from the beginning and checking if any successfully completed runs in the PWC have matching steps.  This is made possible by the retention of data from each stage of the workflows in S3, which can be made available to subsequent workflows.  An example would be if a user wanted to re-run a flood model for the same location, but a different precipitation input, or different model parameters.  The workflow would be able to re-use the querying and pre-processing steps of the non-precipitation data, or even re-use all the querying and pre-processing and just run the model with different parameters.  An example where this provides significant computational savings is in iterative calibration runs where only one or two modules need to be re-run for each iteration (instead of all five) in a flood model example.

\begin{figure*}
    \centering
    \includegraphics[width=1.0\textwidth]{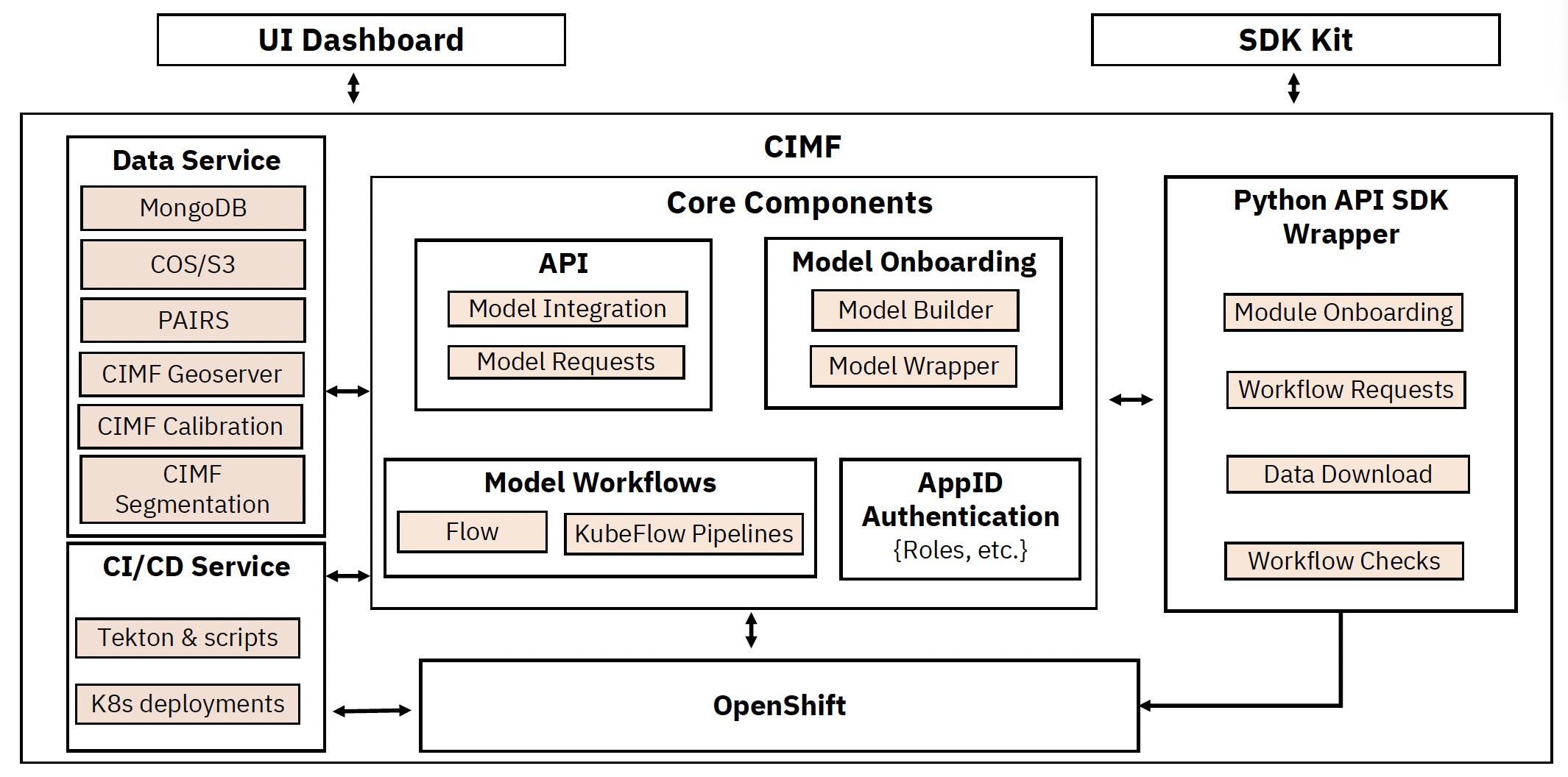}
    \caption{CIMF architecture with core components and default framework microservices.}
    \label{fig:cimf-architecture}
\end{figure*}

Figure~\ref{fig:cimf-architecture} shows the a high-level overview of the framework architecture.  The framework is a microservices-based distributed application running on Red Hat OpenShift.  The core platform consists of a REST API, model and workflow library, model continuous integration / continuous deployment (CI/CD) and workflow orchestrators.  In addition to the core functionality of CIMF, the framework also enables the simple construction of specialized apps/services which can be easily integrated to provide additional capabilities.  These services have their own API for user interaction, but can deploy their workflows and compute to the main CIMF engine (through the API).  Services deployed thus far include those for validating and calibrating models, as well as for hydrological spatial domain segmentation for flood model parallelization.  Once these services are deployed, they can also be referenced by or integrated within CIMF workflows.  An example of this is described in the Experiments section below.  The framework and associated services can be deployed through an infrastructure-as-code package using Kustomize, which allows rapid deployment to OpenShift clusters.  In the future, federation of compute (potentially heterogeneous) will be a priority to support the use of accelerators for scaling and efficient compute for different model types.

\section{Experiments}
In this section, we will present a range of experiments to showcase a number of different types of workflows that CIMF can currently run, although with the flexible and modular nature of our framework any other spatio-temporal geospatial workflow could be onboarded. For these experiments, we used the Integrated Flood Model developed by IBM research~\cite{IFMpaper}. IFM is a hydrological model composed of two main components, a soil and an overland routing model, that aims at providing scalable high-resolution pluvial flood risk forecasting.

\begin{figure}
    \centering
    \includegraphics[width=0.45\textwidth]{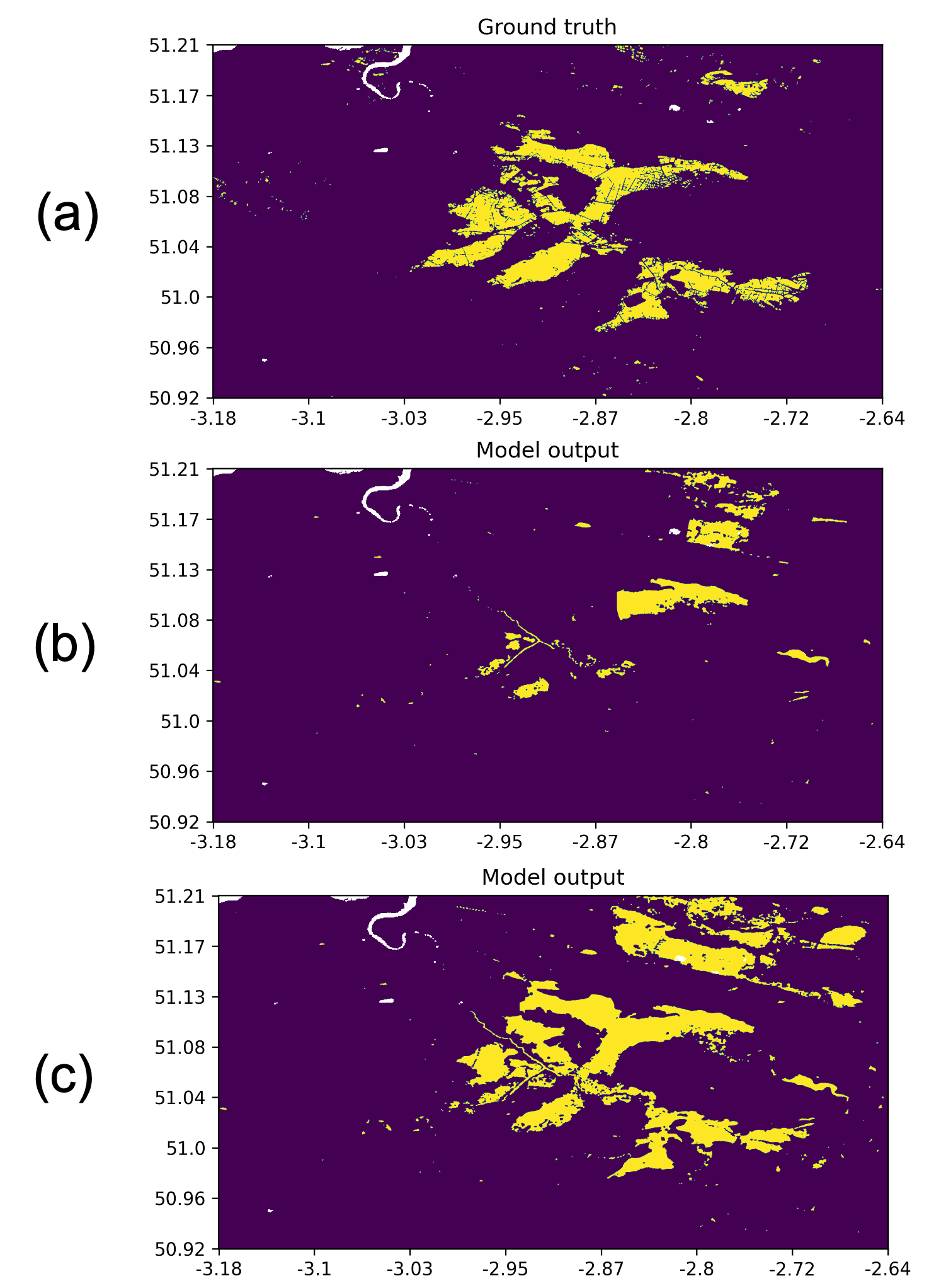}
    \caption{Results from running a calibration workflow for IFM: (a) ground truth; (b) IFM results with initial parameters; (c) IFM results after calibrated parameters}
    \label{fig:cimf-calibration-output}
\end{figure}

Figure~\ref{fig:cimf-calibration-output} shows the results of running a calibration of IFM for an event in the south of the UK in February 2014. The workflow is composed of 100 calibration iterations, where CIMF explores the parameters space within bounds provided by users to maximise Intersection over Union (IoU) between the model output and ground truth indicated by the user, which in this case was obtained from the Copernicus Emergency Management Service (Figure~\ref{fig:cimf-calibration-output}a). As previously mentioned, using an integrated CIMF service the ground-truth flood map can be dynamically generated from remote-sensing data~\cite{ValModule} using an onboarded AI model, with an example that we will be shown in the next step. Figure~\ref{fig:cimf-calibration-output}b shows the model output at the beginning of the calibration (i.e. using the model default parameters), that obtained an IoU of 0.15, with IoU that raised to 0.46 at the end of the calibration (Figure~\ref{fig:cimf-calibration-output}c). This result was possible by submitting a simple json payload to the CIMF API (see Figure~\ref{fig:cimf-payload}), a process which will be made even easier through a python wrapper or UI dashboard.

\begin{figure}
    \centering
    \includegraphics[width=0.45\textwidth]{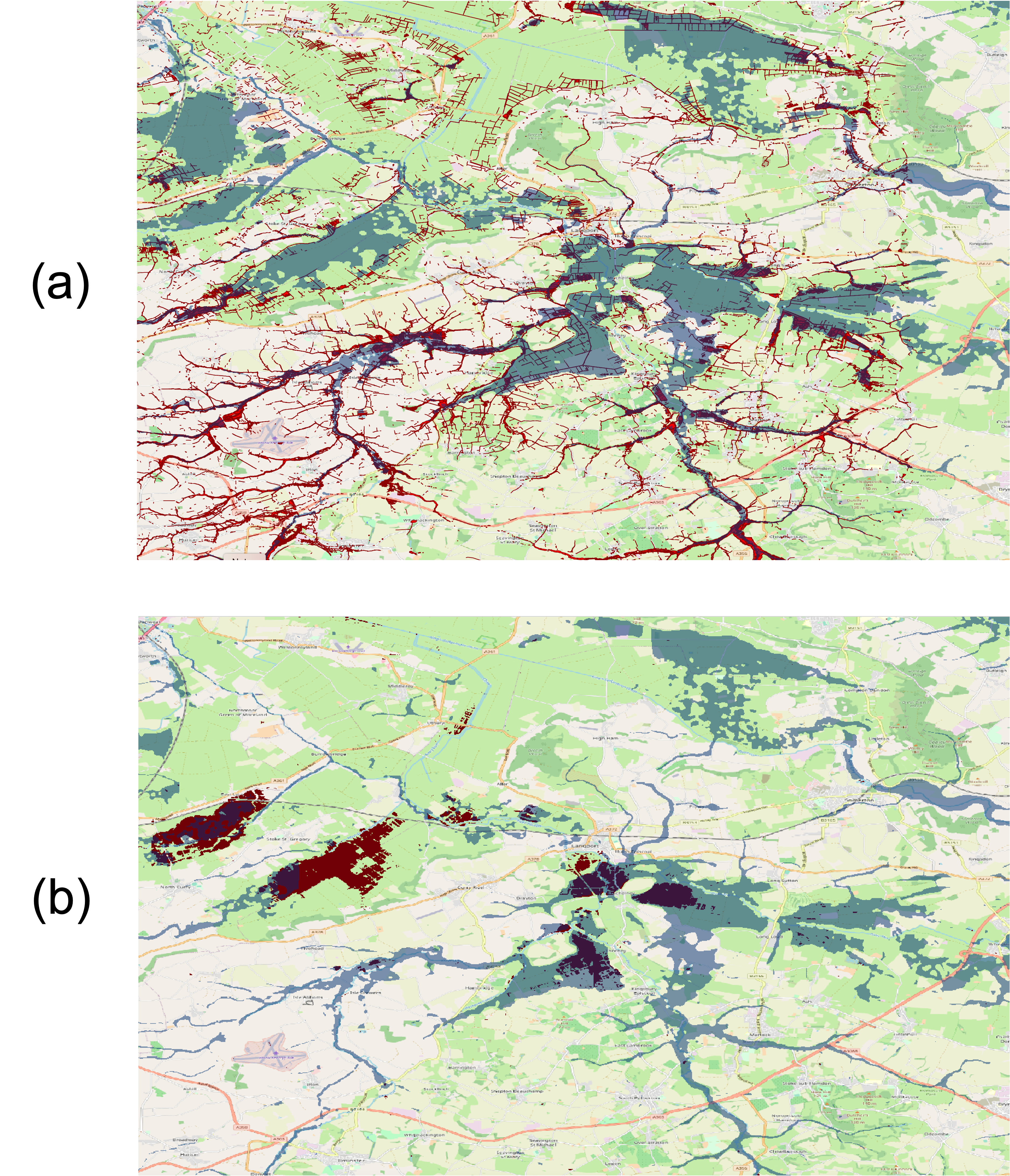}
    \caption{Results from: (a) climatology for the month of December (2001-2021), with red representing flood risk from Environment Agency and in blue the flood risk from the climatology using the calibrated model; (b) sub-seasonal forecast for the last two weeks of December 2021 (blue) compared to AI detected flood on the 29/12/2021 (red)}
    \label{fig:cimf-flood-outputs}
\end{figure}

Once the model is calibrated for the specific area of interest, it is saved and used for further analysis, for example for carrying out a historical climatology or a sub-seasonal forecast. Figure~\ref{fig:cimf-flood-outputs} shows results from these types of analysis for the same area in the South of England. Figure~\ref{fig:cimf-flood-outputs}a shows the flood risk extent obtained by a climatology analysis for the month of December, with an ensemble of IFM models run by CIMF for each December over the 20 years (i.e. a different model run per each year between 2001-2021) using the HadUK-GRID precipitation data and a 30-meter resolution Digital Elevation Model. The CIMF metrics module is subsequently run to calculate relevant flood risk metrics (e.g. probability of flood over a specific threshold, number of days above a specific threshold, and max flood depth), with the blue polygons showing areas where the climatology analysis identified risk of flood above 15 centimeters. This is compared with pluvial flood hazard (i.e. associated to 1 in 100 precipitation events) from the UK Environment Agency. It is possible to observe that there are areas of overlap, but also some differences. These are most likely due to the fact that the Environment Agency data refer to the absolute risk throughout the year, while the climatology focuses only on the month of December. There are also areas where the climatology found larger areas affected by flood risk (e.g. in the fields in the middle of the figure). This is probably due to the resolution of the elevation data used to drive the model not capturing the drainage channels (e.g. visible in the Environment Agency data), which enable water to flow away. The same calibrated model was also used to run a sub-seasonal forecast for the last two weeks of December 2021 with 10 ensemble members, using the same 30 meters resolution and an ECMWF 15-day sub-seasonal forecast for the precipitation data.  Figure~\ref{fig:cimf-flood-outputs}b shows the pluvial flood risk (i.e. above 15 centimeters) extent map obtained with this analysis, again using the CIMF metrics module. This is compared to AI detected flooding on the 29th December 2021 by using the CIMF validation module~\cite{ValModule}. The Figure shows that the areas with flood risk overlap with those that actually flooded. An exception is shown by one of the large fields that flooded on the left of Figure~\ref{fig:cimf-flood-outputs}b, where the flood risk disappears abruptly. After further inspection we could observe that there was indeed registered flood across the ensemble members but it did not reach the 15 centimeters threshold. Again the climatology and sub-seasonal forecast analyses where both ran interacting by submitting simple JSON payloads to the CIMF API. 

\section{Conclusion and Future work }

The Climate Impact Modeling Framework (CIMF) presented in this paper takes a process, which has been traditionally bespoke, fragmented, slow and technical, and provides an ecosystem and a methodology for transforming the way that impact modelling is carried out.  By adopting this modular approach to creating modelling pipelines, CIMF can accelerate the scientific process, both in terms of developing, scaling, and operationalizing impact models, but also in terms of fostering and enabling collaboration, as well as providing different stakeholders with access to a wider, more powerful set of tools.  We have demonstrated example applications of the framework for running flood models in different configurations, benefiting from the composability and scalability nature of the framework.  Work is already ongoing to onboard a range of climate impact and other geospatial models and AI-enabled tools to CIMF.

CIMF is under continual development, including work to streamline the processes around dynamic definition of workflows, the on-boarding of more models and development of more specialised services, and additional support for AI model training through integration with tools and environments such as MLFlow, Ray, and PyTorch.  In addition to this, we are continuing to develop our interfaces, in terms of API wrappers, SDKs, UI and interfaces with a wider range of data sources and external services.  As we develop the framework, we will also build a community of users and content (models, modules, workflows) creators which will expand the framework capabilities further. 

\begin{acks}
This work was supported by the Hartree National Centre for Digital Innovation, a collaboration between the Science and Technologies facilities Council (STFC) and IBM that is funded by the UK Research and Innovation (UKRI) funding agency.

\end{acks}

\bibliographystyle{ACM-Reference-Format}
\bibliography{cimf-paper.bib}

\end{document}